\documentstyle[12pt,frascatiphys,epsfig]{article}
\newcommand{\dgr}{^{\rm o}}
\begin{document}
\title{ 
THE $J^{PC}=0^{++}$ SCALAR MESON NONET AND GLUEBALL\\
OF LOWEST MASS
}
\author{
Peter Minkowski\\
{\em Institute for Theoretical Physics, 
           Univ. of Bern,  CH - 3012 Bern, Switzerland} \\
Wolfgang Ochs   \\
{\em Max Planck Institut f\"ur Physik, D - 80805 Munich, Germany}
}
\maketitle
\baselineskip=14.5pt
\begin{abstract}
The evidence for the low mass $J^{PC}=0^{++}$ states is reconsidered.
We suggest classifying the isoscalars $f_0(980)$ and $f_0(1500)$
as members of the $0^{++}$ nonet, with a mixing rather similar
to that of the pseudoscalars $\eta'$ and $\eta$. The broad state
called $f_0(400-1200)$ or ``sigma'' and the state $f_0(1370)$ are 
considered as different signals from a single broad resonance, 
which we take to be the lowest-lying $0^{++}$ glueball. 
The main arguments in favor of these  hypotheses are presented and 
compared  with theoretical expectations. 
\end{abstract}
\baselineskip=17pt
\section{Introduction}
This session of the workshop is devoted to the study of the ``sigma''
particle, which is related to the large $S$-wave $\pi\pi$
scattering amplitude; it peaks around 800 MeV and again near 1300 MeV. 
The nature of this S wave enhancement was under discussion since the very
beginning of $\pi\pi$ interaction studies\footnote{For a summary of the
early phase of studies in the seventies, see\cite{mms}.} 
and the interpretation is still developing. Its role in S-matrix 
and Regge theory, chiral theories and
 $q\overline q$ spectroscopy is considered since; 
after the advent of
QCD the possibility of glueball spectroscopy\cite{HFPM} has opened up
as well which is in the focus of our attention. In order to obtain the proper
interpretation of the ``sigma'', 
a classification of all low
lying $J^{PC}=0^{++}$ states into the $q\overline q$ nonet and 
glueball states appears necessary. To this end we first discuss the
evidence for the low mass scalar states ($\leq$ 1600 MeV) 
and then proceed with an attempt of their
classification as quarkonium or glueball states
from their properties in production and decay.  We will argue that the
``sigma'' is actually the lightest glueball. 
The main arguments for our classifications will be presented,
further details of this study 
can be found in the recent publication.\cite{mo}
\section{Evidence for light $0^{++}$ states with $I=0$}
The Particle Data Group\cite{PDG} lists the following $I=0$ scalar states:
$f_0(400-1200)$ which is related to the ``sigma'', $f_0(980)$, $f_0(1370)$
and $f_0(1500)$, not all being firmly established. The existence of a
resonance is not only signaled by a peak in the mass spectrum but it
requires in addition that the scattering amplitude moves along 
a full circle in the complex plane (``Argand diagram'').

The first two states have been studied in detail in the phase shift analysis
of elastic $\pi^+\pi^-$ scattering. As discussed by K. Rybicki\cite{Rybicki}, 
the results from high statistics experiments with unpolarized%
\cite{CERN-Munich} and polarized target\cite{CKM} 
have led to an almost unique solution up to 1400 MeV
out of the total of four.
On the other hand, recent data on 
the $\pi^0\pi^0$ final state from GAMS\cite{GAMS} show a different
behaviour of the S-D wave phase  differences above 1200 MeV. A complete phase
shift analysis would provide an important
consistency check with the previous $\pi^+\pi^-$ results. Another experiment
on $\pi^0\pi^0$ pair production 
is in progress (BNL--E852\cite{gunter}), the preliminary 
mass spectrum is shown in fig.1a. 
\begin{figure}[t]
\begin{center}
\mbox{\epsfig{file=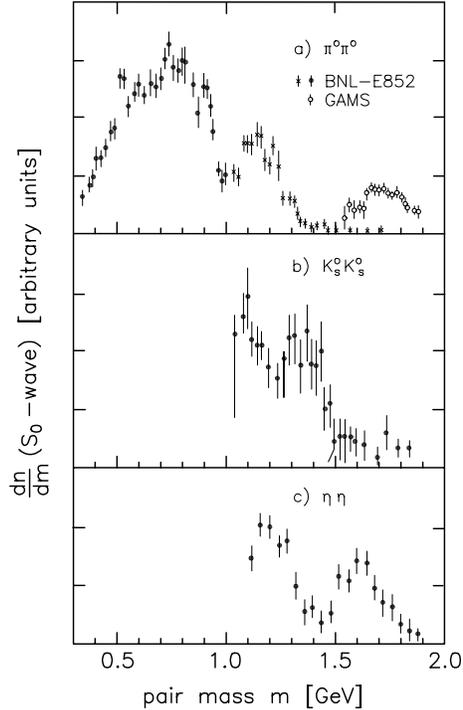,width=60mm}}
\end{center} 
\caption{\it
Isoscalar $S$-wave components of the mass spectra of pseudoscalar pairs
produced in $\pi p$-collisions at small momentum transfers $t$,
(a) $\pi^0\pi^0$ spectrum, recent results\protect\cite{gunter,aps}; 
(b) $K^0_s K^0_s$ spectrum\protect\cite{bnl}
and (c) $\eta\eta$ spectrum.\protect\cite{IIL1}
\label{gbfig3} }
\end{figure} 
One can see a broad spectrum with two or three peaks
(which we refer to as the ``red dragon''). There is no question
about the existence of $f_0(980)$ which causes the first dip near 1 GeV
by its interference with the smooth ``background''.

More controversial is the interpretation of the second peak which appears
in the region 1200-1400 MeV in different experiments. If we remove the
$f_0(980)$ from a global resonance fit of the spectrum the remaining
amplitude phase shift moves slowly through $90\dgr$ near 1000 MeV and
continues rising up to 1400 MeV where it has largely completed 
a full resonance
circle (see also\cite{mp}). A local Breit-Wigner
approximation to these phase shifts yields
\begin{equation}
\textrm{``sigma'':} \qquad\qquad  m\ \sim \ 1000\ \textrm{MeV},
\qquad \Gamma \ \sim \ 1000\ \textrm{MeV}.  \qquad 
\label{sigma}
\end{equation}
In this interpretation the second peak does not correspond to a second
resonance -- $f_0(1370)$ -- but is another signal from the broad object.
A second resonance would require a second circle which is not seen.%
\cite{CERN-Munich,CKM}
Therefore, a complete phase shift 
analysis of the $\pi^0\pi^0$ data in terms of resonances is important
for consolidation.
 
We also investigated whether the state $f_0(1370)$, instead,  appears with
sizable coupling in the inelastic channels 
$\pi\pi \to  K\overline K,\ \eta\eta$ where peaks in the considered
mass region occur as well, although not all at the same position, see
fig.1b,c. To this end we constructed the Argand diagrams for these
channels in fig.2. A similar result for $K\overline K$ has been found 
already from earlier data.\cite{argonne} 

\begin{figure}[t]
\begin{center}
\mbox{\epsfig{file=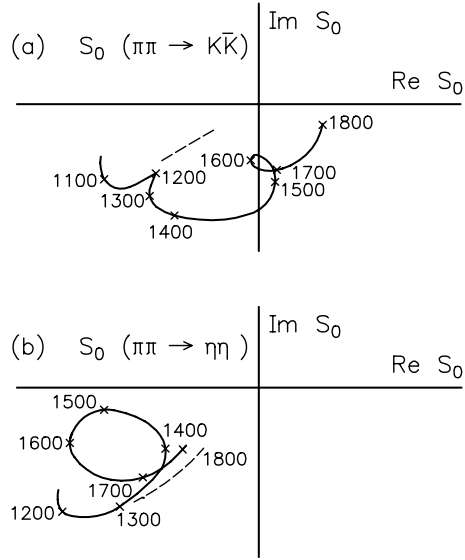,width=60mm}}
\end{center}
\caption{\it
Argand diagrams of the isoscalar $S$-wave amplitudes 
constructed\protect\cite{mo}
from data on the mass spectra shown in fig.1
and the relative phases between $S$ and $D$-waves,\protect\cite{bnl,IIL1} 
assuming a Breit-Wigner
form for the latter.
The numbers indicate the pair masses in MeV, the dashed curves give an
estimate of the background.
\label{gbfig4} }
\end{figure}

The movement of the amplitudes in the complex plane (fig.2) can be interpreted  
in terms of a superposition of a
resonance and a slowly varying background. We identify the
circles with the
$f_0(1500)$ state which has been studied in great detail
by Crystal Barrel.\cite{CB} This resonance can be seen to interfere with
opposite sign in the two channels in figs.2a,b with the background and this
also explains the shift of the peak positions in fig.1b,c.
Thus, the structures in the 1300 MeV region do not correspond to 
additional circles, therefore no additional Breit-Wigner resonance 
$f_0(1370)$ is associated with the respective peaks.
\section{The $J^{PC}=0^{++}$ nonet of lowest mass}
As members of the nonet we take the two isoscalars $f_0(980)$ and 
$f_0(1500)$ which are mixtures of flavor singlet and octet states. 
Furthermore we include the isovector $a_0(980)$ and the strange 
$K^*(1430)$. 
Then the only scalar states with mass below $\sim$
1600 MeV left out up to now are the broad ``sigma'' to which we
come back later and the $a_0(1450)$, which could be a radially excited
state.   

We find the mixing of the $f_0$ states like the one of the
pseudoscalars, namely, with flavour amplitudes
$(u\overline u,d\overline d,s\overline s)$, approximately as
\begin{equation}          
  \label{mixing}           
  \begin{array}{l}        
  \begin{array}{llllll}      
  f_0(980) &\leftrightarrow & \eta^{\prime} (958) \ & \sim &
\frac{1}{\sqrt{6}}(1,\ 1,\ 2) & \quad \textrm{(near}\
\textrm{singlet)} \vspace{2mm}\\
  f_0(1500)& \leftrightarrow & \eta (547) \ & \sim &
\frac{1}{\sqrt{3}}(1,\ 1,\ -1) & \quad \textrm{(near}\ \textrm{octet)}
  \end{array}             
  \end{array}             
\end{equation}            
We have been lead to this
classification and mixing by a number of observations:

\noindent {\it 1. $J/\psi\to \omega,\varphi +X$ decays}\\ 
The branching ratios of  $J/\psi$ into $\varphi \ \eta^{ \prime}  (958)$ and
$\varphi \ f_{ 0}  (980)$ are of similar size and about twice as large as
$\omega \ \eta^{ \prime}  (958)$ and  $\omega \ f_{ 0}  (980)$ which
is reproduced by the above flavor composition.

\newpage
\noindent {\it 2. Gell-Mann-Okubo mass formula}\\
This formula predicts the mass of the octet member $f_0^{(8)}$.
With our octet members $a_0$ and $K^*_0$ as input 
one finds 
$m(f_0^{(8)})=1550$ MeV, or, with the $\eta$-$\eta^{'}$ type mixing included
$m(f_0^{(8)})=1600$ MeV. 
The small deviation of $\sim$10\% in $m^2$ from the mass of
the $f_0(1500)$ is tolerable and can be 
attributed to strange quark mass effects.

\noindent{\it 3. Two body decays of scalars}\\
Given the flavor composition eq.(\ref{mixing}) we can derive the 
decay amplitudes
into pairs of pseudoscalars whereby we allow
for a $s\overline s$ relative amplitude $S$
(for a similar analysis, see\cite{as1}).
In particular, the branching ratios 
\begin{equation}
f_0(980)\to \pi\pi, K\overline K;\quad\ f_0(1500)\to \pi\pi, K\overline K,
\eta\eta, \eta\eta^{'};\quad\ a_0(980),f_0(980)\to \gamma\gamma
\nonumber
\end{equation}
are found  in satisfactory agreement with the data for values
$S$ around 0.5.

\noindent {\it 4. Relative signs of decay amplitudes}\\
A striking prediction is the relative sign of the decay amplitudes of the
$f_0(1500)$ into pairs of pseudoscalars: because of the negative sign in the
$s\overline s$ component, see eq.(\ref{mixing}), the sign of the 
$K\overline K$  decay amplitude is negative with respect to $\eta\eta$ 
decay and
also to the respective $f_2(1270)$ and glueball decay amplitudes. 
This prediction is indeed confirmed by the amplitudes in fig.2a,b which show
 circles pointing in upward and downward directions, respectively. 
If $f_0(1500)$ were a glueball, then both circles should 
have positive sign as in fig.2b, but the experimental results are rather
orthogonal to such an expectation.    

Further tests of our classification 
are provided by the predictions on the decays
$J/\psi\to\varphi/\omega+f_0(1500)$ and the $\gamma\gamma$ decay modes of the
scalars.  

\section{The lightest $0^{++}$ glueball}
In the previous analysis we have classified the scalar mesons in the PDG
tables below 1600 MeV with the exception of $f_0(400-1200)$ and also of
$f_0(1370)$ which we did not accept as standard Breit-Wigner resonance.
We consider the broad spectrum in fig.1a with its two or three peaks as
a single very broad object which interferes with the $f_0$ resonances.
This ``background'' with slowly moving phase appears also in the inelastic
channels (see fig.2). It is our hypothesis that this very broad object 
with parameters eq.(\ref{sigma}) is
the lightest glueball. 
We do not exclude some mixing with the scalar nonet
states but it should be sufficiently small such as to preserve the main
characteristics outlined before. 
We discuss next, how this glueball  assignment fits
with phenomenological expectations.\cite{Close,mo}

\newpage
\noindent {\it 1. The large width}\\
The unique feature of this state is its large width. There are two 
qualitative arguments\cite{mo}  
why this is natural for a light glueball:\\
a) For a heavy glueball one expects a small width as the perturbative
analysis involves a small coupling constant $\alpha_s$ at high masses
(``gluonic Zweig rule\cite{HFPM}). For a light glueball around 1 GeV this
argument doesn't hold any more and a large $\alpha_s$ could yield a large
width.\\
b) The  light $0^{++}$ states are coupled mainly to pairs of
pseudoscalar particles. Then, for a scattering process through a $0^{++}$
channel the external particles are in an S-wave state;
an intermediate $q\overline q$ resonance will be in a P-wave state but an
intermediate $gg$ system in an S-wave again. Therefore the 
overlap of wave functions in the glueball case is larger and 
we expect
\begin{equation}
\Gamma_{gb_0}\ \gg  \ \Gamma_{q\overline q-hadron}. \label{gammaglu}
\end{equation}
\noindent {\it 2. Reactions favorable for glueball production}\\
a) The ``red dragon'' shows up also in the centrally produced systems
in high energy $pp$ collisions\cite{central}  
which are dominated by double Pomeron exchange,
with new results presented by A. Kirk.\cite{Kirk} Because of the gluonic
nature of the Pomeron, this strong production coincides with the
expectations.\\
b) The broad low mass $\pi\pi$ spectrum is also observed in decays
of radially excited states
$\psi'\to\psi(\pi\pi)_s$ and $Y',Y''\to Y(\pi\pi)_s$ which are expected to
be mediated by gluonic exchanges.\\
c) The hadrons in the decay 
$J/\psi \to \gamma+\textrm{hadrons}$ are expected to be produced
through 2-gluon intermediate states which could form
a scalar glueball.
However, in the low mass region
$m<$ 1 GeV only little S-wave in the $\pi\pi$ channel is observed.
 
\noindent {\it 3. Flavour properties}\\
The branching ratios of the $f_0(1370)$ --
which we consider as part of the glueball --
into $K\overline K$ and $\eta\eta$ compare favorably with expectations.

\noindent {\it 4. Suppression in $\gamma\gamma$ collisions}\\
If the mixing of the glueball with charged particles is small it should be
weakly produced in $\gamma\gamma$ collisions.
In the process $\gamma\gamma\to \pi^0\pi^0$ there is 
a dominant peak related to $f_2(1270)$ but, 
in comparison, a very small cross section
in the low mass region around 600 MeV. This could be partially due to 
hadronic rescattering and absorption, partly due to the smallness of the
2 photon coupling of the intermediate states. Unfortunately, the data in the
$f_2$ region leave a large uncertainty on the S-wave fraction 
($<19$\%\cite{Cball}). In a fit to the data which takes into account the
one-pion-exchange Born terms and $\pi\pi$ rescattering the two photon width
of the states $f_2(1270)$ and $f_0(400-1200)$ have been determined\cite{BP}
as 2.84$\pm$0.35 and 3.8$\pm$ 1.5 keV, respectively.
If the $f_0$ were a light quark state like the $f_2$ we might expect
comparable ratios of $\gamma\gamma$ and $\pi\pi$ decay widths, 
but we find (in units of $10^{-6}$)
\begin{equation}
\displaystyle
R_2= \frac{\Gamma(f_2(1270)\to\gamma\gamma)}{\Gamma(f_2(1270)\to\pi\pi)}\ 
  \sim \  15 ;\quad 
R_0= \frac{\Gamma(f_0(400-1200)\to\gamma\gamma)}
       {\Gamma(f_0(400-1200)\to\pi\pi)}\ \sim \ 4-6, 
\label{R02}
\end{equation}
thus, for the scalar state, this ratio is 3-4 times smaller, and it could be
smaller by another factor 3 at about the 2$\sigma$ level.\footnote{
We thank Mike Pennington for the discussions about their analysis.}
A more precise measurement of the S-wave cross section in the $f_2$ region
would be very important for this discussion.

At present, we conclude
that the $2\gamma$ width of the scalar state is indeed surprisingly small.
In this model\cite{BP} 
an intermediate glueball would couple to photons through
the intermediate  $\pi^+\pi^-$ channel. 

\noindent {\it 5. Quark-antiquark and gluonic components in $\pi\pi$
scattering}\\
In the dual Regge picture the $2\to 2$  scattering amplitude is built
either from the sequence of s-channel resonances
or from the sequence of t-channel Regge poles. There is a second component 
(``two component duality''\cite{fh}) which corresponds to the Pomeron in the
t-channel and is dual to a ``background'' in the direct s-channel.
If the Pomeron is related to glueballs, then one should have,
by crossing, a third component with a glueball in the direct
s-channel, dual to exotic exchange.\cite{mo}

\begin{figure}[t]
\begin{center}
\mbox{\epsfig{file=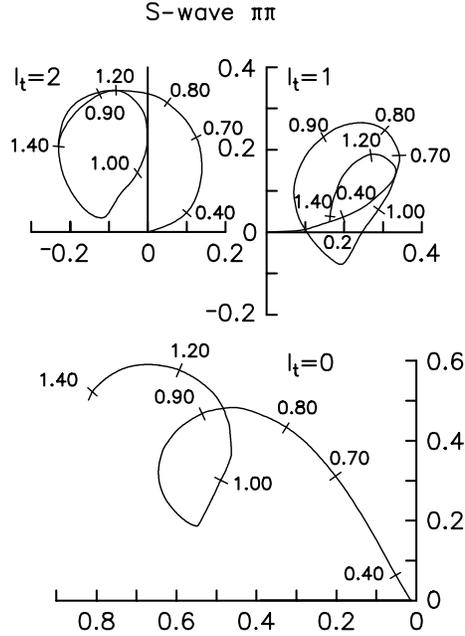,width=60mm}}
\end{center}
\caption{\it
Amplitudes for the $\pi\pi$ S-wave for definite t-channel Isospin $I_t$
determined by Quigg\protect\cite{quigg}. The $I_t=0$ amplitude 
has a large background which is not related to the $q\overline q$
resonances in the dual model approach.
\label{gbfig5} }
\end{figure}

The existence of the ``background'' process can be demonstrated by
constructing the amplitudes for definite t-channel isospin $I_t$. Such an
analysis has been carried out by Quigg\cite{quigg} 
for $\pi\pi$ scattering
 and is shown in fig.3. Similar to what has been found in $\pi N$
scattering\cite{hz}
there are essentially background-free resonance circles for $I_t\neq 0$, but 
in the $I_t=0$ amplitude (Pomeron exchange) 
the background rises with energy and
is sizable already below 1 GeV. We take this result as a further hint
that low energy $\pi\pi$ scattering is not dominated by 
$q\overline q$ resonances alone.

\section{Theoretical expectations}
\subsection{QCD results on glueballs}
\noindent {\it 1. Lattice QCD}\\
 In the calculation without sea-quarks (``quenched approximation'')
one finds the lightest glueball in the $0^{++}$ channel at masses 1500-1700
MeV (recent review\cite{Teper}). These results have motivated various
recent searches and scenarios for
the lightest glueball. The identification
with the well established $f_0(1500)$ state, either with or without mixing
with other states, has some phenomenological
difficulties, especially the negative amplitude sign into $K\overline
K$ (fig.2a).

Some changes of these QCD predictions 
may occur if the full unquenched calculation is carried out. The first
results by Bali et al.\cite{lattunq} indicate a decrease of the glueball
mass with the quark masses; the latter are still rather large 
and correspond to $m_{\pi}\sim 700 \ldots 1000 $ MeV.
For the moment we conclude that our 
light glueball hypothesis is not necessarily in conflict with the lattice
QCD results.
 
\noindent {\it 2. QCD sum rules}\\
The saturation of the sum rules for the $0^{++}$ glueball was found 
impossible with a single state near 1500 MeV alone in a recent
analysis.\cite{Nar} Rather, the inclusion of a light glueball component
was required and assumed to be coupled to
states $\sigma_B(1000)$ and $\sigma_{B'}(1370)$. Already before, a 
sum rule solution with a light glueball $\sim 500$ MeV was proposed.%
\cite{bagan}  

\noindent {\it 3. Bag model}\\
In a model which consideres quarks and gluons
to be confined in a bag of comparable size
and with radiative QCD corrections included,\cite{Barnes} 
the lightest glueball was suggested for $0^{++}$ at around 1 GeV mass.

\subsection{Scalar nonet and effective Sigma variables}
An important precondition for the assignment of glueball states is the
understanding of the low mass $q\overline q$ spectroscopy.

\noindent {\it 1. Renormalizable linear sigma models}\\
These models realize the spontaneous chiral symmetry breakdown and
represent an attractive theoretical approach to the scalar and
pseudoscalar mesons. An example is the  approach by T\"{o}rnqvist%
\cite{Torn} which starts from a ``bare'' nonet respecting the OZI rule 
while the observed hadron spectrum is strongly distorted by
unitarization effects.

In an  alternative approach,\cite{njl} %
one starts from a 3-flavor Nambu-Jona-Lasinio
model but includes a renormalizable  effective action for the sigma fields
with an instanton induced axial $U(1)$ symmetry-breaking term 
along the suggestion by t'Hooft.\cite{thooft}
In this model
 $f_0(1500)$
is near the octet and the light isoscalar  near the singlet state;
different options are pursued\cite{njl} for 
$f_0(980)$ and $a_0(980)$, at least one of them 
should be a non-$q\overline q$ state. 
This suggestion of a large singlet-octet mixing and the classification of
the $f_0(1500)$ is close to our phenomenological findings in sect.3.

\noindent {\it 2. General effective QCD potential}\\
In our approach\cite{mo} we do not restrict ourselves to renormalizable
interaction terms.
In this way the
consequences of chiral symmetry in different limits for the quark masses can
be explored in a general QCD framework.\cite{PM} In particular, it is
possible to keep both $f_0(980)$ and $a_0(980)$ as $q\overline q$ states.
Their degeneracy in mass can be obtained, although not predicted.
An expansion to first order in the strange quark mass is investigated.
The Gell-Mann-Okubo formula is obtained in this approximation; 
with an $\eta$-$\eta'$ type mixing the observed states  discussed
in sect.3, with  $f_0(1500)$ as the heaviest
member of the nonet near the octet state, can be realized.

\section{Conclusions}
We found a classification of the low lying $J^{PC}=0^{++}$ states
which explains a large body of experimental and phenomenological results.
The $q\overline q$ nonet includes  $f_0(980)$ and $f_0(1500)$ with
mixing similar to the pseudoscalars $\eta'$ and $\eta$, furthermore 
$a_0(980)$ and $K^*(1430)$;  $\eta'$ and $f_0(980)$ appear as
genuine parity doublet.

The lightest glueball is identified with the broad ``sigma'' corresponding
to $f_0(400-1200)$ and  $f_0(1370)$ of the PDG. The basic triplet of
light binary glueballs is completed\cite{mo} by the states $\eta(1440)$ with
$0^{-+}$ and $f_J(1710)$ with $2^{++}$, not discussed here.   

It will be important to further study production and decay of the states
under discussion. Some particular questions we came across here include:
a) unique phase shift solution for $\pi\pi$ scattering above 1 GeV
for both charge modes ($+-$ and $00$), b)
production of $f_0(1500)$ in $J/\psi$ decays, c) S-waves in radiative 
$J/\psi$ decays and d) $\gamma\gamma$ widths of the scalar particles.

It remains an open question in this approach,
 though, what  the physical origin
of the $a_0 - f_0$ mass degeneracy is and where
the mirror symmetry of the mass patterns
in the scalar and pseudoscalar nonets comes from.
A possible explanation for the latter  
structure is suggested by a renormalizable model with an instanton induced
$U_A(1)$-breaking interaction.

\end{document}